# Sex-by-age differences in the resting-state brain connectivity

Sean D. Conrin[1*], Liang Zhan[2*], Zachery D. Morrissey[1], Mengqi Xing[1,3],
Angus Forbes[4], Pauline Maki[1], Mohammed R. Milad[1], Olusola Ajilore[1], Alex D. Leow[1,3]

1. Department of Psychiatry, University of Illinois at Chicago, IL, USA
2. Computer Engineering Program, University of Wisconsin-Stout, WI, USA
3. Department of Bioengineering, University of Illinois at Chicago, IL, USA
4. Department of Computational Media, University of California, Santa Cruz, CA, USA

* indicate equal contribution.

**Abstract**

Connectomics is a state-of-the-art framework that models brain structure and function interconnectivity as a network in order to better understand the human brain as a whole, rather than narrowly focusing on select regions-of-interest. MRI-derived connectomes can be structural, usually based on diffusion-weighted MR imaging, or functional, usually formed by examining fMRI blood-oxygen-level-dependent (BOLD) signal correlations. Recently, we developed a novel method for assessing the hierarchical modularity of functional brain networks—the probability associated community estimation (PACE). PACE is unique in that it permits a dual formulation, thus yielding equivalent connectome modular structure regardless of whether one considers positive or negative edges. This method was rigorously validated using the 1,000 functional connectomes or F1000 project data set (F1000, RRID:SCR_005361) (Biswal, Mennes et al. 2010) and the Human Connectome Project (HCP, RRID:SCR_006942) (Van Essen, Smith et al. 2013) and we detected novel sex differences in resting-state connectivity that were not previously reported.(Zhan, Jenkins et al. 2017) This current study more thoroughly examined sex differences as a function of age and their clinical correlates, with findings supporting *a basal configuration framework*. To this end, we found that men and women do not significantly differ in the 22-25-year-old age range. However, these same non-significant differences attained statistical significance in the 26-30 age group (p=0.003), while becoming highly statistically significant in the 31-35 age group (p<0.001). At the most global level, areas of diverging sex difference include parts of the prefrontal cortex and the temporal lobe, amygdala, hippocampus, inferior parietal lobule, posterior cingulate, and precuneus. Further, we identified statistically different self-reported summary scores of inattention, hyperactivity, and anxiety problems between men and women. These self-reports additionally divergently interact with age and the *basal configuration* between sexes. In sum, our study supports a paradigm change in how we conceptualize the functional connectome, shifting away from simple concepts such as the default mode network (DMN), and towards thinking globally and probabilistically how the brain exhibits dynamic sex-specific connectivity configuration as a function of age, and the role this sex-by-age configuration at rest (i.e., the *basal configuration*) might play in mental health frequency and presentation, including symptom patterns in depression.

## 1. Introduction

In efforts to better understand the human connectome, various approaches have been used to identify and measure the modularity of brain connectivity. In these efforts the brain is generally divided into a collection of communities or "modules". Frequently, these modules can be sub-divided into submodules, which then demonstrate hierarchical modularity and *near decomposability* (the autonomy of modules from one another). Modules and sub-modules are comprised of a series of nodes with tight interconnectivity, whereas nodes of differing modules have lesser connectivity. The connections between nodes are referred to as edges and can be either positive or negative in fMRI connectomics. A positive edge indicates that the activity in one node is positively correlated with that in the connected node, whereas a negative edge indicates the presence of an inverse relationship between the two. When these concepts are applied to fMRI-derived networks, network organization identifies functionally related or "coupled" regions.

The complexity and volume of data associated with large networks is enormous and thus much work has been done to develop algorithms that better characterize and measure modularity. In these various processes there are a number of areas in which methods differ in their approach. As a popular approach, maximizing Q modularity metric is an NP-hard problem and typically speed of computation comes with a compromise. For instance, the very efficient fast unfolding method does not guarantee global optimization. (Vincent, Jean-Loup et al. 2008) Another area of variations is how positive and negative edges are accounted for. More frequently, the focus of computation is based on the recognition and measure of positive edges as this is fitting for many types of networks. However, in measuring brain connectivity through fMRI networks, which usually focuses on fMRI blood-oxygen-level-dependent (BOLD) signal correlations, there is the presence of both positive and negative edges.

Most published studies have employed variable approaches of ignoring, thresholding, binarizing, or arbitrary down-weighting to account for these negative edges. Although quite different in their approaches, the common similarity is that the data involving negative edges is in some degree not rigorously accounted for.

To better address the assessment of negative edges, we recently developed and published a novel method for assessing the modularity of functional brain networks—the probability associated community estimation (PACE). Most importantly, PACE permits a dual formulation, thus yielding equivalent connectome modular structure regardless of whether one considers positive or negative edges, by exploiting how frequent BOLD signal correlation between two regions is negative (the 'edge negativity'). This method was rigorously validated using resting-state fMRI data from the 1,000 functional connectomes or F1000 project data set (F1000, RRID:SCR_005361) (Biswal, Mennes et al. 2010) and the Human Connectome Project (HCP, RRID:SCR_006942) (Van Essen, Smith et al. 2013) and we demonstrated that negative correlations alone are sufficient in understanding resting-state connectome modularity.

As we progressed with this method, we explored whether our approach might be useful in the study of sex-based differences in healthy brain function, with the understanding that this might contribute to the discussion of how and why men and women difference in their expression of mental illness. When compared to various existing Q maximization based formulations applied to the same two data sets, PACE yielded results that are both, consistent with existing methods, yet more stable and reproducible than alternative methods. Moreover, as a result of its superior reproducibility (and thus robustness), PACE was able to detect novel subtle sex differences in resting-state connectivity that were not previously reported with Q-based methods.

These areas of sex difference we reported are traditionally considered to be either part of, or closely related to brain areas thought to be responsible for self-referential thinking, including part of the prefrontal cortex (PFC) and the temporal lobe, amygdala, hippocampus, precuneus, and inferior parietal lobule. Along this line, conceptually at the most global level resting state networks can be generally classified as either "task-positive" or TPN or "task-negative" or TNN, with the latter (TNN) concept encompassing the default mode network (DMN) of the brain that is more narrowly defined to be core regions activated with self-referential thinking. (Greicius, Krasnow et al. 2003)

In this study, we further explored whether our approach might be useful in the study of sex-based differences in healthy brain function, with the understanding that this might contribute to the discussion of how and why men and women differ in their expression of mental illness.

To this end, we examined sex differences in resting-state connectome as a function of age, with the hypothesis that differences may emerge and become more prominent as age progresses. We comprehensively explored how these differences relate to image resolution as it regards to the validity of these differences. Finally, we investigated how globally the segregation, integration and interaction between TPN and TNN during the resting state may relate to self-reports of common psychopathology traits in terms of the sexes. As this investigation unfolded, it became apparent that a significant portion of the discussion regarding this matter ought to be directed towards whether our current conceptualization of the default mode network is overly simplistic and possibly contributes to an artificial division from other portions of the brain.

## 2. Methodology

### 2.1 Data

The data we used in this study is 811 subjects' resting state fMRI connectome data from the Human Connectome Project (released in December 2015, named as HCP900 Parcellation+Timeseries+Netmats, https://db.humanconnectome.org/data/projects/HCP_900). Three different spatial dimensions of brain networks are explored: 100x100, 200x200, and 300x300, all derived using independent component analysis or ICA. For details of the dataset and the procedure for connectome construction, please refer to HCP's official website and respective references.(Sporns, Tononi et al. 2005, Wedeen, Hagmann et al. 2005, Burgel, Amunts et al. 2006, Hagmann, Kurant et al. 2007, Chiang, Barysheva et al. 2009) The study subjects' demographics are shown in **Table 1.**

**Table 1.** Participant demographics

| Age | Male (Age in years) | Female (Age in years) | ASR Depressive Problems Raw Score mean ± std (range) | ASR Anxisty Problems Raw Score mean ± std (range) | ASR Inattention Problems Raw Score mean ± std (range) | ASR Hyperactivity Problems Raw Score mean ± std (range) |
| --- | --- | --- | --- | --- | --- | --- |
| 22~25 | n= 106 (23.45 ± 1.08) | n= 70 (23.66 ± 1.11) | 4.32±3.60 (0~19) | 4.05±2.62 (0~11) | 3.54 ± 2.44 (0~11) | 2.80 ± 2.17 (0~10) |
| 26~30 | n= 152 (27.91 ± 1.37) | n= 197 (28.09 ± 1.48) | 4.28±3.74 (0~22) | 3.79±2.77 (0~14) | 3.03 ± 2.27 (0~14) | 2.54 ± 2.04 (0~11) |
| 31~35 | n= 106 (32.44 ± 1.24) | n= 180 (32.86 ± 1.37) | 3.62±2.86 (0~15) | 3.61±2.43 (0~12) | 2.93 ± 2.32 (0~11) | 2.24 ± 1.90 (0~8) |

## 2.2 Community estimation

In this study, we adopted the probability associated community estimation (PACE) (Zhan, Jenkins et al. 2017) framework to extract the hierarchical modularity of the resting-state functional connectome. Rather than working with the magnitude of BOLD signal correlations, PACE operates on the edge positivity/negativity probability pair ($P^+/P^-$) such that edges most frequently anti-correlated should be placed across communities (if two regions are always anti-activating, the edge positivity = 0 and negativity = 1). Algorithmically, PACE identifies a partition of the set of edges $V$ (into N communities) that maximizes the PACE benefit function $\Psi$, defined as the difference between mean inter-community edge negativity and mean intra-community edge negativity. Moreover, considering the duality between $P^+$ and $P^-$ (($P^+ + P^-$ =1 for all edges) PACE optimization problem thus permits an equivalent dual form. Formally,

$$\underset{C^1 \cup C^2 \cup \ldots \cup C^N = V, C^i \cap C^j = \emptyset \text{ for all } i \neq j}{\operatorname{argmax}} \left\{ \frac{\sum_{1 \leq n < m \leq N} \overline{P^-(C^n, C^m)}}{N(N-1)/2} - \frac{\sum_{1 \leq n \leq N} \overline{P^-(C^n)}}{N} \right\} =$$

$$\underset{C^1 \cup C^2 \cup \ldots \cup C^N = V, C^i \cap C^j = \emptyset \text{ for all } i \neq j}{\operatorname{argmax}} \left\{ \frac{\sum_{1 \leq n \leq N} \overline{P^+(C^n)}}{N} - \frac{\sum_{1 \leq n < m \leq N} \overline{P^+(C^n, C^m)}}{N(N-1)/2} \right\}$$

Please refer to (Zhan, Jenkins et al. 2017) for details of PACE including implementation and verification.

## 2.3 Constructing the PACE null model and testing the statistical significance of each bifurcation

In our current implementation, N is set to 2 and PACE attempts, for each branch at a specific PACE level, to further split that branch into 2 subsequent groups. Then, a nonparametric procedure can be followed to determine the level of statistical significance for such a split. By stopping a branch from further splitting when there is evidence against it, PACE can in theory, yield any number of communities (no longer restricted to powers of 2).

Here, we propose a comprehensive procedure that constructs the null distribution given the observed data. Slightly Improving upon the original procedure proposed in [1], here instead of permuting $P^+/P^-$ within each edge (original procedure) we sample the null distribution (i.e., assuming there is no modular patterns of co-/anti- activation) of the PACE benefit function $\Psi$ by randomly choosing two edges and exchanging their edge negativity/positivity probability pairs (randomization is iterated over the entire connectome). Then, this entire process is repeated for 1000 times, yielding 1000 samples of $\Psi$ under null. Last, to determine the significance of each split, the actual $\Psi$ achieved by the original data is compared to the 1000 sampled $\Psi$ values under null at the same PACE level; if the former lies within the top 5% of the latter, such a split is determined to be significant ($P < 0.05$). (**Figure 1**)

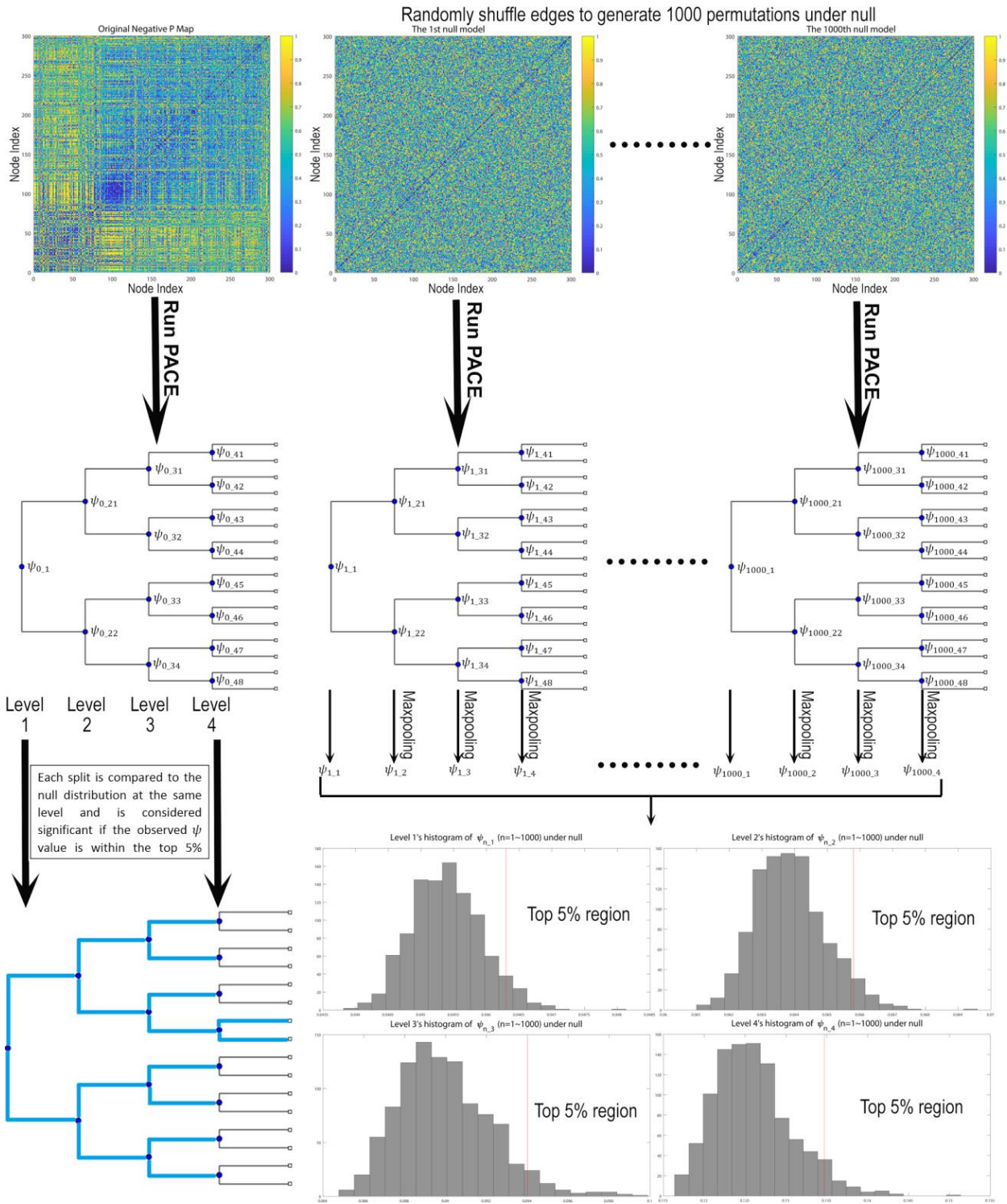

**Figure 1.** Testing the statistical significance of each bifurcation for the PACE hierarchical modularity tree. First we derive the community structure from the observed data (the original edge negativity P map). Then we randomly shuffle edges to create 1000 permutations under null and for each permutation, we run PACE to obtain bifurcation trees (and conduct max-pooling for level 2 and up), yielding the null distribution ψ for each level. Then the bifurcation of the original data at each split will be treated as significant only when the observed ψ is within top 5% with respect to the null distribution.

## 2.4 Clinical correlates of connectome modularity: relate systems-level PACE modular structure to subject-level characteristics

First, as part of the HCP data release, we obtained the Adult Self-Report (ASR) DSM Depressive, Anxiety, Inattention, and Hyperactivity Problems scores (**Table I**), for each subject and tested if there are sex differences after controlling for age. In particular, using a general linear model incorporating an intercept, main effects, as well as a sex-age interaction term, the statistical significance of a sex effect is tested by centering age within the age range 22-35 across all subjects, for all ASR scores.

Separately, for each of the three resolutions available for HCP (100-ROI, 200-ROI, 300-ROI) we applied PACE to extract modularity (separately for each sex as well as for the combined total sample) and determined the optimal level of bifurcation using the null-model procedure introduced above, thus at the most global level (i.e. 1st level of PACE) yielding two modules, operationally defined as the task negative network (TNN) that includes regions traditionally considered DMN, and the task positive network (TPN; defined as the other network that does not include DMN regions). To summarize the overall activity of TNN and TPN, for each individual we computed the average correlation value within TNN (avg-TNN) and TPN (avg-TPN).

To explore behavioral correlates of TPN and TNN, we conducted partial correlations, controlling for age, between avg-TNN and avg-TPN and each of the four ASR DSM Problems scores.

## 3. Results
### 3.1 ASR DSM Problems Scores

For ASR DSM Inattention and Hyperactivity Problems scores men are significantly higher (assessed at mean age of 28.75 years: for Inattention Problems men higher than women by 0.36 points, standard error SE=0.17, p=0.031; for Hyperactivity Problems men higher than women by 0.33 points, SE=0.026, p=0.0023), while women's self-reported ASR Anxiety Problems scores are significantly higher (assessed at mean age of 28.75 years: women higher than men by 0.83 points, SE= 0.19, p=8.5e-06). In addition, overall there are age effects for both Anxiety and Hyperactivity Problems scores (lower scores as age progresses: for Anxiety Problems beta= -

0.107, SE=0.034, p=0.0017; for Hyperactivity Problems beta= -0.080, SE=0.026, p=0.0023) but not for Inattention Problems score.

For ASR DSM Depressive Problems, there is a significant age effect (beta = -0.137, SE=0.045, p=0.0026) but not a significant sex difference.

### 3.2 PACE Modularity Results

Across the entire sample, PACE-derived modularity at the most global level (yielding two modules operationally defined as the TPN, in red, and the TNN, in green) is shown in **Figure 2** for each sex and each of 3 parcellation resolutions (100-, 200-, and 300- ROIs). As expected, visually sex differences are clearer for the higher resolutions (P value of sex-differences = 0.0001 for 100 ROIs, and <0.0001 for both 200 and 300 ROIs).

Second, **Figure 3** further illustrates modularity sex differences using the 300-ROI resolution at the global level while **Figure 4** details the corresponding complete hierarchical modularity, visualized as bifurcation trees, after applying our null-distribution procedure (see section 2.3). Interestingly, we identified an additional split in men during this procedure (resulting in 8 modules for women and 9 for men).

### 3.3 PACE Modularity as a function of age

Next, we explore connectome modularity as a function of age, with results shown in **Figure 5** where we visualize the sex-specific modularity in each of three age groups (22~25, 26~30 and 31~35). Although sex-differences did not reach statistical significance in the 22-25 year old age group, a trend difference is visually observed in several areas. This trend then reaches statistical significance in the 26-30 year old age group (p=0.003) and becomes highly statistically significant in the 31-35 age group (p<0.001). Interestingly, visually female modularity remains largely consistent across the three age groups (a head-to-head comparison between the 22-25 y/o group and the 31-35 y/o group was indeed statistically not significant), whereas males exhibit visually notable changes across age, particularly from the 26-30 to 31-35 age groups. Also areas of significant changes are in brain regions where in men "transitions" occur, as they age, gradually from more probabilistically task-negative (green) into task-positive (red) (including the precuneus, the inferior parietal lobule, the prefrontal cortex, the hippocampus, the amygdala, and the middle temporal gyrus).

## 3.4 Secondary analyses: correlation between PACE modularity and ASR DSM Problems Scores

Secondary post-hoc partial correlation analyses (controlling for age) confirmed that avg-TPN is negatively correlated with ASR Anxiety and Inattention Problems scores in men, but not in women; all other correlations are statistically non-significant (r= -0.131 and p=0.01 for Anxiety Problems, r= -0.127 and p=0.015 for Inattention Problems; p values uncorrected).

Last, across the entire sample, avg-TPN also positively correlates with avg-TNN, suggesting a synergistic relationship between them than antagonistic (r=0.37 and p=1.4e-28, **Figure 6**).

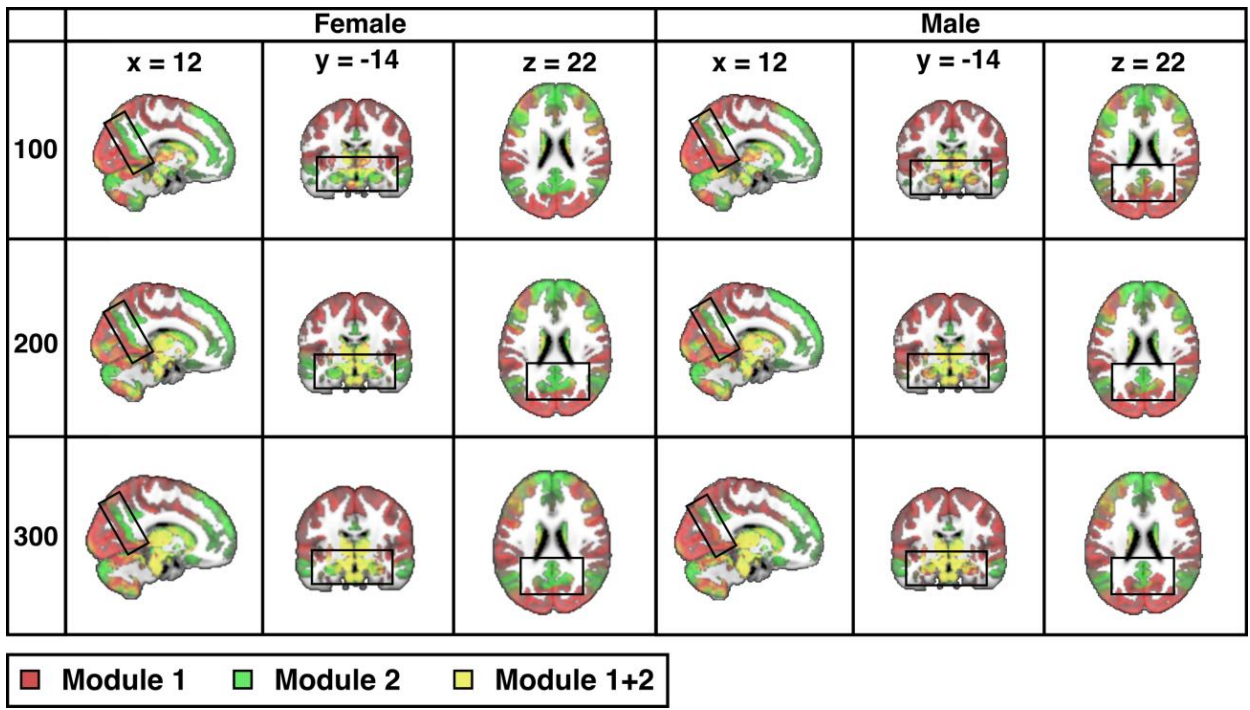

**Figure 2**. Sex differences in resting-state modularity revealed using the probability associated community estimation (PACE). Slices obtained from 100-ROI, 200-ROI, and 300-ROI resolutions using the Human Connectome Project (HCP) data. At PACE Level 1, two brain modules are extracted, here shown as the red community (corresponding to task positive network or TPN) and the green community (corresponding to task negative community or TNN). Mixing of communities is shown in the overlay in yellow (since HCP parcellation is ICA-based, components may overlap thus resulting in the mixing of TPN and TNN). As expected, with increasing spatial resolution (from 100 to 300 ROIs), sex differences also become more significant (for both 200-ROI and 300-ROI resolution p < 0.0001). Differences between the sexes (boxed) include precuneus, hippocampus, and amygdala. MNI coordinates: x=12, y=-14, z=22.

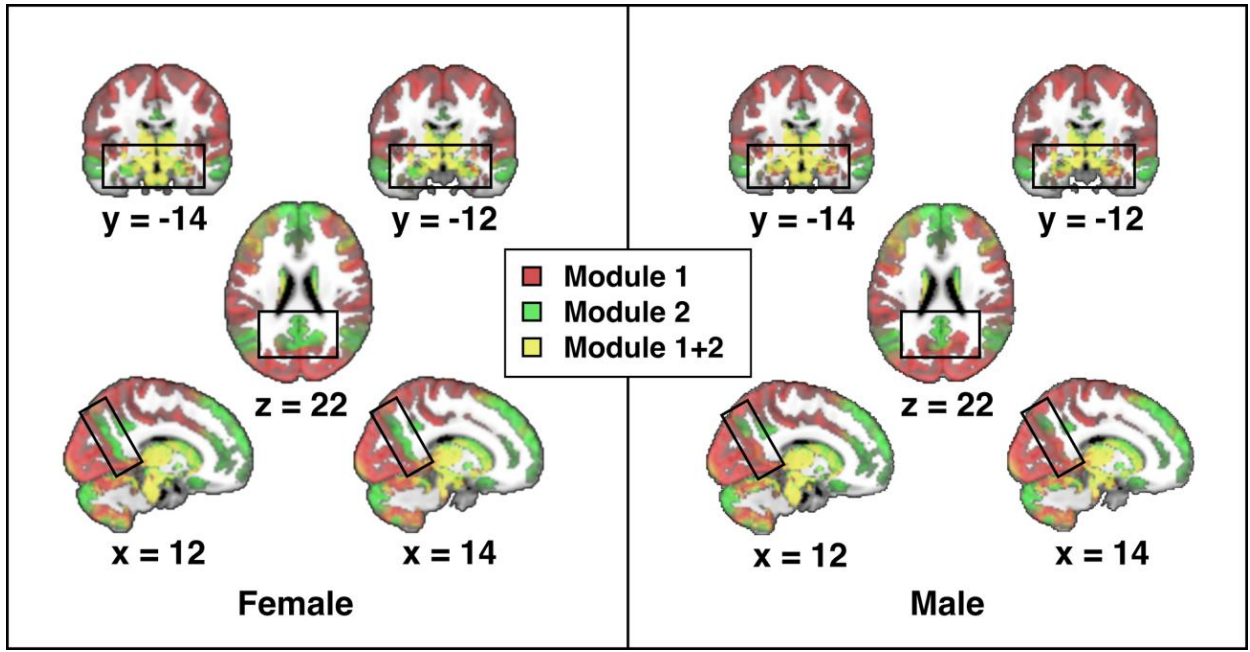

**Figure 3.** More detailed sex differences of global modularity (PACE Level 1) for the 300-ROI HCP data (M: 364, F: 447), in which two modules are extracted: the red community (task-positive) and the green community (task-negative). Areas of intersection between modules are indicated in yellow. Boxed areas indicate regions of significant differences across sex ($p < 0.0001$), including precuneus, hippocampus, and amygdala. MNI coordinates: x=12,14; y=-14, -12; z=22.

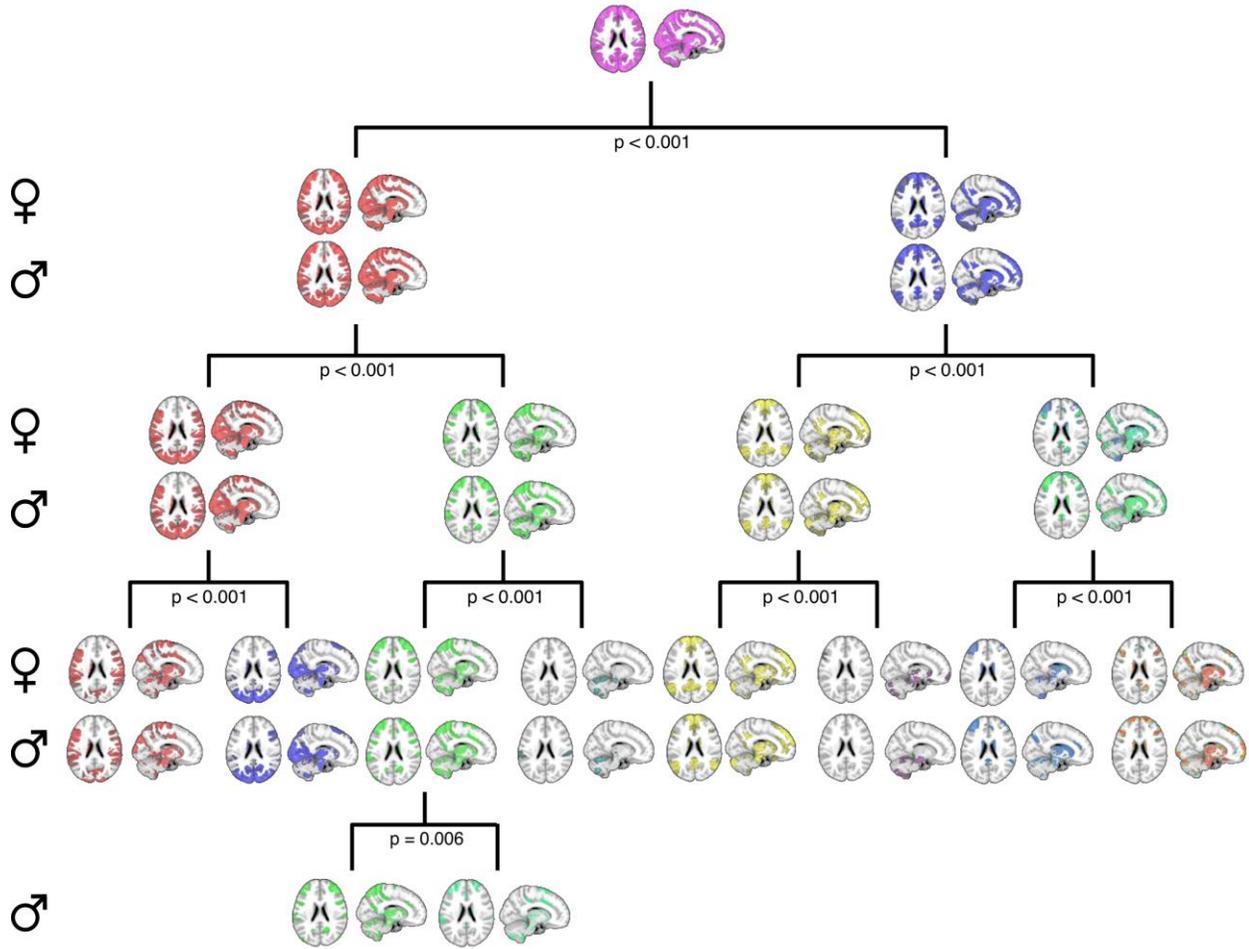

**Figure 4.** A graphical representation of the PACE-derived male and female hierarchical modular structure, optimally determined using the proposed null-model procedure. Female communities are displayed in the top row of each level with their male counterparts displayed on the bottom row. Axial and sagittal slices are shown for each community. Note that while female community bifurcation was no longer significant after PACE Level 3 (eight total communities), a further significant bifurcation was observed in males (third community from left at PACE level 3; p=0.006), yielding nine total communities. MNI coordinates: x=12, z=22.

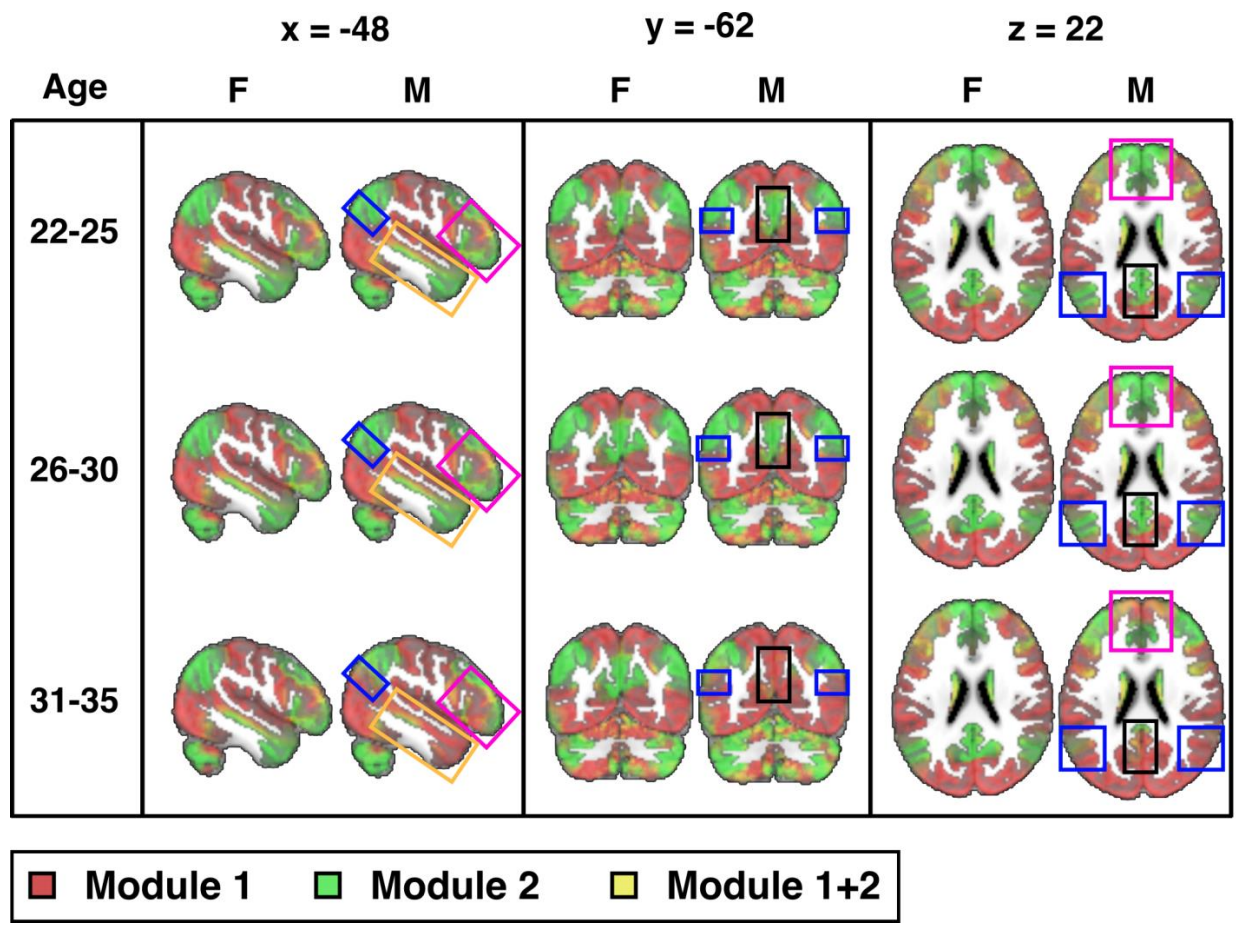

**Figure 5.** Sex differences across three age groups (visualized at PACE Level 1): 22-25 (M: 106, F: 70), 26-30 (M: 152, F: 197), and 31-35 (M: 106, F: 180) years-old (cf. **Table 1**) from the 300-ROI HCP data. Sex differences were not statistically significant in the 22-25-year-old group, but reached statistical significance in the 26-30-year-old group (p=0.003) and becomes highly statistically significant in the 31-35 group (p<0.001). Interestingly, visually female systems-level global modularity remains largely consistent across the three age groups, whereas males exhibit visually notable changes across age, particularly from the 26-30 to 31-35 age groups. Areas of significant changes are in brain regions where transitions occur from TNN (green) to TPN (red) in men but not in women (boxed areas: black: precuneus; blue: inferior parietal lobule; magenta: prefrontal cortex; orange: middle temporal gyrus). F=female; M=male. MNI coordinates: x=-48, y=-62, z=22.

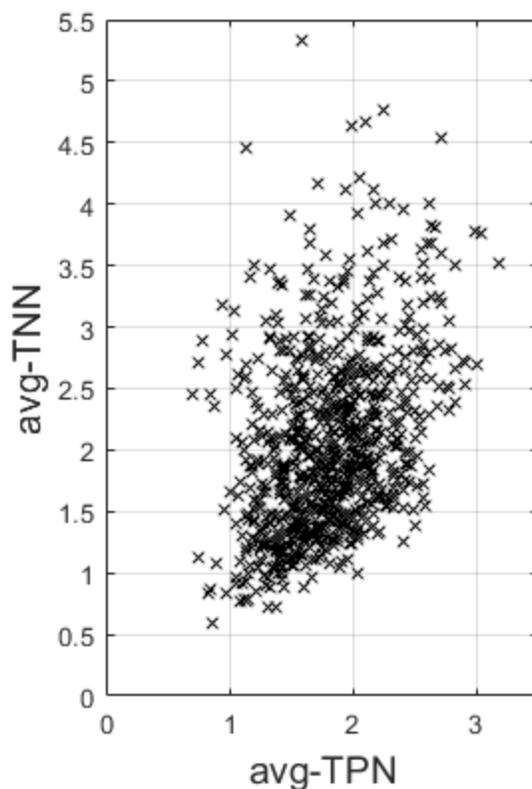

**Figure 6.** This figure plots the average Fisher's z-transformed correlation within TNN (avg-TNN) against that computed within TPN (avg-TPN) across the entire sample. There is a highly significant positive correlation (r=0.37 and p=1.4e-28) between the two, suggesting a synergistic relationship between them than antagonistic.

## 4. Discussion

Recently, we developed PACE as a novel method for assessing the hierarchical modularity of resting-state functional brain networks. PACE is unique in that it permits a dual formulation, thus yielding invariant connectome modular structures regardless of whether one considers positive or negative edges. This method was rigorously validated using the F1000 project and the HCP dataset, and we demonstrated that PACE results are generally consistent with those from existing methods, while more stable and reproducible. As a result, PACE was able to detect novel subtle sex differences in resting-state connectivity that were not previously reported.

In this study, we further comprehensively examined these novel connectome sex differences over the course of early adulthood, in particular how they are modulated by age as well as their clinical correlates. Our findings lead to a more nuanced picture elucidating: 1) how fMRI-derived brain connectivity exhibits both sex-specific configuration and age-dependent dynamic re-configuration, the latter primarily in men, from early twenties to mid-thirties, and 2) how this sex-by-age configuration during the resting state (the *basal configuration*) relates to self-reports of common psychopathology traits in otherwise healthy subjects. We call this novel and more nuanced framework the *basal configuration framework*.

Although sex-differences were not identified at a level of statistical significance in the 22-25-year-old age group, a trend difference is observed in several areas in the brain. This trend then reaches statistical significance in the 26-30 year old age group ($p=0.003$) and becomes highly statistically significant in the 31-35 age group ($p<0.001$). Specifically, in areas either designated as or functionally closely coupled with DMN we identified a gradually shifting connectome configuration over time in men (vs. women), away from probabilistically more "task negative" and towards probabilistically more "task positive". Thus, in contrast to conventional binary notions of, e.g., DMN being "task-negative", the basal configuration framework provides a much more flexible approach to the resting state, by considering the connectome-wide probabilistic distribution or "configuration" of inter-regional connectivities. In this sense, basal configuration's probabilistic framework frees up a brain region from having to be hard-parsed (e.g., DMN vs non-DMN), thus allowing functional brain units (i.e., modules) to work synergistically at times, and to act nearly-decomposable at other times. Of interest, at the same time that basal configuration differences are identified with increasing statistical significance with age, the differences also show increasing statistical significance within individual age groups with increasing image resolution.

What might contribute to the sex differences observed in the present study? It is well documented in the literature that men and women differ in the activation of the amygdala, hippocampus, and other prefrontal regions in response to emotional stimuli. (Maeng and Milad 2015) Some studies have shown that hormones like estrogen and progesterone might modulate the resting-state functional connectivity within various nodes of the DMN. (Abu-Zeitone, Peterson et al. 2014, Engman, Linnman et al. 2016) These data thus suggest that some of the sex differences in this current study may be due to differences in sex hormones between men and women, and/or hormonal fluctuations in women. Interestingly, several recent studies additionally showed that exogenous hormones in the form of oral contraceptives impact functional activation to emotional stimuli and resting state functional connectivity (Abu-Zeitone, Peterson et al. 2014,

Hwang, Zsido et al. 2015, Petersen and Cahill 2015) and women using oral contraceptives have smaller hippocampal volume compared to women not using contraceptives.(Hertel, Konig et al. 2017) Stress also impacts men and women differently.(Merz, Wolf et al. 2013), with oral contraceptive users having significant and constant elevation of cortisol and altered hypothalamic-pituitary axis (HPA), both of which have been shown to impact functional activation of many nodes tested here.(Hertel, Konig et al. 2017) In sum, it remains unclear to us why the sex differences would become more pronounced with age, and thus future studies ought to focus on examining how fluctuations of gonadal hormones and use of OCs might contribute to some of the differences noted here.

Given these sex-based age-dependent basal configuration differences in otherwise healthy people, a natural further course of inquiry is to ask how these differences in healthy functioning might relate to the known differences in the prevalence and symptom expression of mental illness between the sexes. Among clear differences between men and women in terms of mental health is that women are much more likely than men to experience episodes of depression.(Nolen-Hoeksema, Morrow et al. 1993, Weissman, Bland et al. 1996, Piccinelli and Wilkinson 2000, Seney and Sibille 2014) This is an important area of exploration as much research has been directed towards identifying explanatory models for these differences without conclusive results. Examples of frequently investigated factors are differences in: hormonal development, societal roles, communication patterns and ways of coping with stressors. However, the lack of definitive conclusions to date suggests that additional and unrecognized factors might be at play. (Piccinelli and Wilkinson 2000) As such, it may be useful to consider whether sex-based differential brain network connectivity might be one of these unrecognized factors. Clearly it is a complex picture and it is important to note that exposure to sex hormones is already known to affect brain network connectivity. One example of this is discussed by Ottowitz et al., that addition of estrogen to post-menopausal women is shown to increase connectivity between the hippocampus and the prefrontal cortex, which as part of the fronto-limbic circuitry known to play important roles in depression.(Ottowitz, Siedlecki et al. 2008) While Maki et al. provide another example in which they find that perimenopausal women exposed earlier to hormone replacement therapy perform better in various cognitive tasks than those who are exposed after a longer time untreated.(Maki, Dennerstein et al. 2011)

One area in which differential brain network connectivity possibly contributes to sex-based differences in depression is rumination. Rumination is the repetitive thinking and focus on negative mood states.(Shors, Millon et al. 2017) It is also a form of self-referential processing of information.(Nejad, Fossati et al. 2013) As summarized in Nejad et al's 2013 review, several areas are identified as playing a role in self-referential processing: medial prefrontal cortex, anterior and posterior cingulate cortex, insula, temporal pole, hippocampus and amygdala.(Nolen-Hoeksema, Morrow et al. 1993, Gusnard, Akbudak et al. 2001, Kelley, Macrae et al. 2002, Fossati, Hevenor et al. 2003, Phan, Wager et al. 2004, Ochsner and Gross 2005, Johnson, Raye et al. 2006, Schmitz and Johnson 2006, van der Meer, Costafreda et al. 2010, Nejad, Fossati et al. 2013) Note that in our current study we identified sex-based differences in areas highly overlapping with these.

Rumination and its role in depression are frequently explored in regard to propensity towards developing depressive episodes, duration and severity of episodes, and also of recovery from depression. It has been shown that women are more likely than men to engage in rumination during depressive episodes, (Nolen-Hoeksema, Morrow et al. 1993) and that those engaged in ruminative responses during depressive episodes are likely to experience longer episodes. (Nolen-Hoeksema, Morrow et al. 1993) As identified previously and also supported by our results, in women the amygdala is probabilistically more strongly coupled with what is conventionally considered the "task-negative network" that also contains other DMN regions. (Zhan, Jenkins et al. 2017) Given these findings, it is plausible that differences in connectivity in areas involved with rumination and the default mode network provide a propensity towards women to "think more about self" at rest and subsequently explain at least a portion of the differences between men and women in terms of their rates of depression and its clinical presentations. Further, this conjecture is supported by previously reported sex differences in functional connectivity (and thus cognitive strategies) during emotion processing/regulation. For example, in Mcrae et al. 2018, men were found to down-regulate activity in the amygdala without corresponding increases in cortical activity, while women reappraising negative emotion are not as likely to do this independently and instead recruit more cortical activity. (Mcrae, Ochsner et al. 2008, Lungu, Potvin et al. 2015)

Another major difference recognized in our study is in regard to self-reported inattention, hyperactivity, and anxiety problems scores. While depression problems scores are similar between the sexes, men as a whole have higher self-reported inattention and hyperactivity scores than women (women by contrast have higher overall anxiety problems scores). Further, correlation analyses revealed that overall avg-TPN is positively correlated with avg-TNN, and in the male group the former further inversely correlated with attention problems and anxiety problems scores. Taken together, our results support that at rest 'task-negative' regions do not necessarily compete against or deactivate 'task-positive' ones, as has been suggested in some previous literature, or further as we argued with our basal configuration framework simple notions of *task-negative* vs. *task-positive* modes do not capture the full dynamic and complex patterns of brain interconnectivity. Rather, we advocate for probabilistically considering brain interconnectivity as a whole, with its modularity exhibiting a temporally dynamic configuration such that brain regions do not exhibit fixed attributes or identities but instead work synergistically and in conjunction with one another in a flexible and dynamic way.

Also, the fact that we only found basal configuration correlations with self-reports in the male group merits more discussion. First, we note that it is less likely due to a lack of power in the female group, since we did find robust correlations in the male group. Guided by recent lines of evidence suggesting that in women cognition is a function of the menstrual cycle due to differential effects of estrogen and progesterone on cognition (Upadhayay and Guragain 2014), we conjecture that in women the basal configuration and self-reports such as ASR scores are also dependent on the menstrual cycle. Thus, it is plausible that the lack of correlations in the female group is merely a consequence of the fact that the functional imaging data and the self-reports were obtained on different days and thus during different phases of the menstrual cycle.

Last, note that our sex-difference findings in self-reports are generally in line with Gur et al.'s 2002 paper in which they found that men have earlier declines in frontotemporal areas associated with attention, inhibition and memory. (Gur and Gur 2002) Indeed, men tend to express symptoms of depression differently than women in terms of acting-out behaviors such as substance abuse, restlessness and suicide. (Piccinelli and Wilkinson 2000, Walinder and Rutzt 2001) While this difference in expression of depressive symptoms is not fully understood, it is possible that the tendency towards higher degrees of inattention and hyperactivity in otherwise healthy men at least partially explains why men tend to exhibit more externalizing behaviors when depressed. This finding is also interesting to consider in the context of Ingalhalikar et al.'s findings regarding sex-based connectivity differences in adolescents in which the development of greater inter-hemispheric connectivity in females leads to better performance in attention tasks.(Ingalhalikar, Smith et al. 2014) As a whole, this supports a concept of sex-based differences in basal configuration leading to different sex-based behavioral tendencies in both healthy and disease-state individuals. Together with our correlation results, these findings suggest that the degree of basal configuration's functional coupling may thus serve as an imaging biomarker for the self-regulation of externalizing behaviors.

## 5. Conclusion

In sum, this study contributes to a growing literature on the limitations of our current conceptualization of resting-state networks. To this end, the notion of the default mode network being the network active at rest, thus driving resting-state brain dynamics, while the activation of other networks is responsible for taking over during active tasks is misleading at best. Rather, we argue that this perspective is an oversimplification of the actual complexity of brain connectivity, as on a subject-level with increasing default mode network activity the functional activity in other networks (generally considered to be "task positive" networks) increases as well. Taken in conjunction with previously published studies in which increasing connectivity is a more important factor in overall functionality,(Smith, Nichols et al. 2015) this study lays the foundation of a more nuanced framework, termed the basal configuration, for better conceptualizing the resting state. To this end, our results supported that: 1) the basal configuration exhibits distinct sex-specific dynamics by mid 30s, 2) the basal configuration diverges during early adulthood between the sexes, in that there is globally an age-modulated reconfiguration between the 'task-positive' and 'task-negative' networks primarily in men but not women, 3) the basal configuration correlates with self-reported measures of personality traits, at least in men, 4) whereas in women the basal configuration is further conjectured to likely exhibit a strong dependence on the menstrual cycle.

# 6. References

Abu-Zeitone, A., D. R. Peterson, B. Polonsky, S. McNitt and A. J. Moss (2014). "Oral contraceptive use and the risk of cardiac events in patients with long QT syndrome." Heart Rhythm **11**(7): 1170-1175.

Biswal, B. B., M. Mennes, X. N. Zuo, S. Gohel, C. Kelly, S. M. Smith, C. F. Beckmann, J. S. Adelstein, R. L. Buckner, S. Colcombe, A. M. Dogonowski, M. Ernst, D. Fair, M. Hampson, M. J. Hoptman, J. S. Hyde, V. J. Kiviniemi, R. Kotter, S. J. Li, C. P. Lin, M. J. Lowe, C. Mackay, D. J. Madden, K. H. Madsen, D. S. Margulies, H. S. Mayberg, K. McMahon, C. S. Monk, S. H. Mostofsky, B. J. Nagel, J. J. Pekar, S. J. Peltier, S. E. Petersen, V. Riedl, S. A. Rombouts, B. Rypma, B. L. Schlaggar, S. Schmidt, R. D. Seidler, G. J. Siegle, C. Sorg, G. J. Teng, J. Veijola, A. Villringer, M. Walter, L. Wang, X. C. Weng, S. Whitfield-Gabrieli, P. Williamson, C. Windischberger, Y. F. Zang, H. Y. Zhang, F. X. Castellanos and M. P. Milham (2010). "Toward discovery science of human brain function." Proc Natl Acad Sci U S A **107**(10): 4734-4739.

Burgel, U., K. Amunts, L. Hoemke, H. Mohlberg, J. M. Gilsbach and K. Zilles (2006). "White matter fiber tracts of the human brain: three-dimensional mapping at microscopic resolution, topography and intersubject variability." Neuroimage **29**(4): 1092-1105.

Chiang, M. C., M. Barysheva, D. W. Shattuck, A. D. Lee, S. K. Madsen, C. Avedissian, A. D. Klunder, A. W. Toga, K. L. McMahon, G. I. de Zubicaray, M. J. Wright, A. Srivastava, N. Balov and P. M. Thompson (2009). "Genetics of brain fiber architecture and intellectual performance." J Neurosci **29**(7): 2212-2224.

Engman, J., C. Linnman, K. R. Van Dijk and M. R. Milad (2016). "Amygdala subnuclei resting-state functional connectivity sex and estrogen differences." Psychoneuroendocrinology **63**: 34-42.

Fossati, P., S. J. Hevenor, S. J. Graham, C. Grady, M. L. Keightley, F. Craik and H. Mayberg (2003). "In search of the emotional self: an fMRI study using positive and negative emotional words." Am J Psychiatry **160**(11): 1938-1945.

Greicius, M. D., B. Krasnow, A. L. Reiss and V. Menon (2003). "Functional connectivity in the resting brain: a network analysis of the default mode hypothesis." Proc Natl Acad Sci U S A **100**(1): 253-258.

Gur, R. E. and R. C. Gur (2002). "Gender differences in aging: cognition, emotions, and neuroimaging studies." Dialogues Clin Neurosci **4**(2): 197-210.

Gusnard, D. A., E. Akbudak, G. L. Shulman and M. E. Raichle (2001). "Medial prefrontal cortex and self-referential mental activity: relation to a default mode of brain function." Proc Natl Acad Sci U S A **98**(7): 4259-4264.

Hagmann, P., M. Kurant, X. Gigandet, P. Thiran, V. J. Wedeen, R. Meuli and J. P. Thiran (2007). "Mapping human whole-brain structural networks with diffusion MRI." PLoS One **2**(7): e597.

Hertel, J., J. Konig, G. Homuth, S. Van der Auwera, K. Wittfeld, M. Pietzner, T. Kacprowski, L. Pfeiffer, A. Kretschmer, M. Waldenberger, G. Kastenmuller, A. Artati, K. Suhre, J. Adamski, S. Langner, U. Volker, H. Volzke, M. Nauck, N. Friedrich and H. J. Grabe (2017). "Evidence for Stress-like Alterations in the HPA-Axis in Women Taking Oral Contraceptives." Sci Rep **7**(1): 14111.

Hwang, M. J., R. G. Zsido, H. Song, E. F. Pace-Schott, K. K. Miller, K. Lebron-Milad, M. F. Marin and M. R. Milad (2015). "Contribution of estradiol levels and hormonal contraceptives to sex differences within the fear network during fear conditioning and extinction." BMC Psychiatry **15**: 295.

Ingalhalikar, M., A. Smith, D. Parker, T. D. Satterthwaite, M. A. Elliott, K. Ruparel, H. Hakonarson, R. E. Gur, R. C. Gur and R. Verma (2014). "Sex differences in the structural connectome of the human brain." Proc Natl Acad Sci U S A **111**(2): 823-828.

Johnson, M. K., C. L. Raye, K. J. Mitchell, S. R. Touryan, E. J. Greene and S. Nolen-Hoeksema (2006). "Dissociating medial frontal and posterior cingulate activity during self-reflection." Soc Cogn Affect Neurosci **1**(1): 56-64.

Kelley, W. M., C. N. Macrae, C. L. Wyland, S. Caglar, S. Inati and T. F. Heatherton (2002). "Finding the self? An event-related fMRI study." J Cogn Neurosci **14**(5): 785-794.

Lungu, O., S. Potvin, A. Tikasz and A. Mendrek (2015). "Sex differences in effective fronto-limbic connectivity during negative emotion processing." Psychoneuroendocrinology **62**: 180-188.

Maeng, L. Y. and M. R. Milad (2015). "Sex differences in anxiety disorders: Interactions between fear, stress, and gonadal hormones." Horm Behav **76**: 106-117.

Maki, P. M., L. Dennerstein, M. Clark, J. Guthrie, P. LaMontagne, D. Fornelli, D. Little, V. W. Henderson and S. M. Resnick (2011). "Perimenopausal use of hormone therapy is associated with enhanced memory and hippocampal function later in life." Brain Res **1379**: 232-243.

Mcrae, K., K. Ochsner, I. Mauss, J. D. Gabrieli and J. J. Gross (2008). "Gender Differences in Emotion Regulation: An fMRI Study of Cognitive Reappraisal." Group Processes and Intergroup Relations **11**(2): 143-162.

Merz, C. J., O. T. Wolf, J. Schweckendiek, T. Klucken, D. Vaitl and R. Stark (2013). "Stress differentially affects fear conditioning in men and women." Psychoneuroendocrinology **38**(11): 2529-2541.

Nejad, A. B., P. Fossati and C. Lemogne (2013). "Self-referential processing, rumination, and cortical midline structures in major depression." Front Hum Neurosci **7**: 666.

Nolen-Hoeksema, S., J. Morrow and B. L. Fredrickson (1993). "Response styles and the duration of episodes of depressed mood." J Abnorm Psychol **102**(1): 20-28.

Ochsner, K. N. and J. J. Gross (2005). "The cognitive control of emotion." Trends Cogn Sci **9**(5): 242-249.

Ottowitz, W. E., K. L. Siedlecki, M. A. Lindquist, D. D. Dougherty, A. J. Fischman and J. E. Hall (2008). "Evaluation of prefrontal-hippocampal effective connectivity following 24 hours of estrogen infusion: an FDG-PET study." Psychoneuroendocrinology **33**(10): 1419-1425.

Petersen, N. and L. Cahill (2015). "Amygdala reactivity to negative stimuli is influenced by oral contraceptive use." Soc Cogn Affect Neurosci **10**(9): 1266-1272.

Phan, K. L., T. D. Wager, S. F. Taylor and I. Liberzon (2004). "Functional neuroimaging studies of human emotions." CNS Spectr **9**(4): 258-266.

Piccinelli, M. and G. Wilkinson (2000). "Gender differences in depression. Critical review." Br J Psychiatry **177**: 486-492.

Schmitz, T. W. and S. C. Johnson (2006). "Self-appraisal decisions evoke dissociated dorsal-ventral aMPFC networks." Neuroimage **30**(3): 1050-1058.

Seney, M. L. and E. Sibille (2014). "Sex differences in mood disorders: perspectives from humans and rodent models." Biol Sex Differ **5**(1): 17.

Shors, T. J., E. M. Millon, H. Y. Chang, R. L. Olson and B. L. Alderman (2017). "Do sex differences in rumination explain sex differences in depression?" J Neurosci Res **95**(1-2): 711-718.

Smith, S. M., T. E. Nichols, D. Vidaurre, A. M. Winkler, T. E. Behrens, M. F. Glasser, K. Ugurbil, D. M. Barch, D. C. Van Essen and K. L. Miller (2015). "A positive-negative mode of population covariation links brain connectivity, demographics and behavior." Nat Neurosci **18**(11): 1565-1567.

Sporns, O., G. Tononi and R. Kotter (2005). "The human connectome: A structural description of the human brain." PLoS Comput Biol **1**(4): e42.

Upadhayay, N. and S. Guragain (2014). "Comparison of cognitive functions between male and female medical students: a pilot study." J Clin Diagn Res **8**(6): BC12-15.

van der Meer, L., S. Costafreda, A. Aleman and A. S. David (2010). "Self-reflection and the brain: a theoretical review and meta-analysis of neuroimaging studies with implications for schizophrenia." Neurosci Biobehav Rev **34**(6): 935-946.

Van Essen, D. C., S. M. Smith, D. M. Barch, T. E. Behrens, E. Yacoub, K. Ugurbil and W. U.-M. H. Consortium (2013). "The WU-Minn Human Connectome Project: an overview." Neuroimage **80**: 62-79.

Vincent, D. B., G. Jean-Loup, L. Renaud and L. Etienne (2008). "Fast unfolding of communities in large networks." Journal of Statistical Mechanics: Theory and Experiment **2008**(10): P10008.

Walinder, J. and W. Rutzt (2001). "Male depression and suicide." Int Clin Psychopharmacol **16 Suppl 2**: S21-24.

Wedeen, V. J., P. Hagmann, W. Y. Tseng, T. G. Reese and R. M. Weisskoff (2005). "Mapping complex tissue architecture with diffusion spectrum magnetic resonance imaging." Magn Reson Med **54**(6): 1377-1386.

Weissman, M. M., R. C. Bland, G. J. Canino, C. Faravelli, S. Greenwald, H. G. Hwu, P. R. Joyce, E. G. Karam, C. K. Lee, J. Lellouch, J. P. Lepine, S. C. Newman, M. Rubio-Stipec, J. E. Wells, P. J. Wickramaratne, H. Wittchen and E. K. Yeh (1996). "Cross-national epidemiology of major depression and bipolar disorder." JAMA **276**(4): 293-299.

Zhan, L., L. M. Jenkins, O. E. Wolfson, J. J. GadElkarim, K. Nocito, P. M. Thompson, O. A. Ajilore, M. K. Chung and A. D. Leow (2017). "The significance of negative correlations in brain connectivity." J Comp Neurol.